\documentclass[twocolumn]{aastex631}

\usepackage{xspace}
\usepackage{CJK}

\hypersetup{
    colorlinks=true,
    linkcolor=blue,
    filecolor=magenta,      
    urlcolor=cyan,
}

\usepackage[normalem]{ulem} 
\usepackage{cancel} 

\usepackage{color}

\newcommand{\kms}{\ensuremath{\rm{km\,s}^{-1}}\xspace}
\newcommand{\alphaco}{\ensuremath{\alpha_{\rm{CO}}}\xspace}

\newcommand{\uJy}{\ensuremath{\mu\rm{Jy}}\xspace}

\newcommand{\lir}{\ensuremath{L_{\rm{IR}}}\xspace}

\newcommand{\Tdust}{\ensuremath{T_{\rm{dust}}}\xspace}

\newcommand{\Mstar}{\ensuremath{M_{\rm{*}}}\xspace}

\newcommand{\Mmol}{\ensuremath{M_{\rm{mol}}}\xspace}
\newcommand{\Mmolgal}{\ensuremath{M_{\rm{mol,gal}}}\xspace}
\newcommand{\Mmoltail}{\ensuremath{M_{\rm{mol,tails}}}\xspace}
\newcommand{\Zstar}{\ensuremath{Z_{\rm{*}}}\xspace}
\newcommand{\Msol}{\ensuremath{\rm{M}_\odot}\xspace}

\newcommand{\Zsol}{\ensuremath{\rm{Z}_\odot}\xspace}

\newcommand{\lpgal}{\ensuremath{\rm{L}_{\rm{CO(2-1),gal}}'}\xspace}
\newcommand{\lptail}{\ensuremath{\rm{L}_{\rm{CO(2-1),tails}}'}\xspace}

\newcommand{\oiii}{[O{\scriptsize III}]\xspace}

\newcommand{\etal}{et~al.\xspace}
\newcommand{\hst}{\textit{HST}\xspace}

\newcommand{\squiggle}{SQuIGG\ensuremath{\vec{L}}E\xspace}

\shortauthors{J.~S.~Spilker, et~al.}
\shorttitle{Molecular Gas in Tidal Tails}

\begin{document}
\begin{CJK*}{UTF8}{gbsn}

\title{Star Formation Suppression by Tidal Removal of Cold Molecular Gas\\from an Intermediate-Redshift Massive Post-starburst Galaxy}

\correspondingauthor{Justin S. Spilker}
\email{jspilker@tamu.edu}

\author[0000-0003-3256-5615]{Justin~S.~Spilker}
\affiliation{Department of Physics and Astronomy and George P. and Cynthia Woods Mitchell Institute for Fundamental Physics and Astronomy, Texas A\&M University, 4242 TAMU, College Station, TX 77843-4242, US}

\author[0000-0002-1714-1905]{Katherine~A.~Suess}
\affiliation{Department of Astronomy and Astrophysics, University of California, Santa Cruz, CA, 95064, USA}
\affiliation{Kavli Institute for Particle Astrophysics and Cosmology and Department of Physics, Stanford University, Stanford, CA 94305, USA}

\author[0000-0003-4075-7393]{David~J.~Setton}
\affiliation{Department of Physics and Astronomy and PITT PACC, University of Pittsburgh, Pittsburgh, PA 15260, USA}

\author[0000-0001-5063-8254]{Rachel~Bezanson}
\affiliation{Department of Physics and Astronomy and PITT PACC, University of Pittsburgh, Pittsburgh, PA 15260, USA}

\author[0000-0002-1109-1919]{Robert~Feldmann}
\affiliation{Institute for Computational Science, University of Zurich, CH-8057, Zurich, Switzerland}

\author[0000-0002-5612-3427]{Jenny~E.~Greene}
\affiliation{Department of Astrophysical Sciences, Princeton University, Princeton, NJ 08544, USA}

\author[0000-0002-7613-9872]{Mariska~Kriek}
\affiliation{Leiden Observatory, Leiden University, NL-2300 RA Leiden, Netherlands}

\author[0000-0003-4422-8595]{Sidney~Lower}
\affiliation{Department of Astronomy, University of Florida, Gainesville, FL 32611, USA}

\author[0000-0002-7064-4309]{Desika~Narayanan}
\affiliation{Department of Astronomy, University of Florida, Gainesville, FL 32611, USA}

\author[0000-0003-1535-4277]{Margaret~Verrico}
\affiliation{Department of Physics and Astronomy and PITT PACC, University of Pittsburgh, Pittsburgh, PA 15260, USA}

\begin{abstract}

Observations and simulations have demonstrated that star formation in galaxies must be actively suppressed to prevent the formation of over-massive galaxies. Galactic outflows driven by stellar feedback or supermassive black hole accretion are often invoked to regulate the amount of cold molecular gas available for future star formation, but may not be the only relevant quenching processes in all galaxies. We present the discovery of vast molecular tidal features extending up to 64\,kpc outside of a massive $z=0.646$ post-starburst galaxy that recently concluded its primary star-forming episode. The tidal tails contain $(1.2\pm0.1)\times10^{10}$\,\Msol of molecular gas, $47\pm5$\% of the total cold gas reservoir of the system. Both the scale and magnitude of the molecular tidal features are unprecedented compared to all known nearby or high-redshift merging systems. We infer that the cold gas was stripped from the host galaxies during the merger, which is most likely responsible for triggering the initial burst phase and the subsequent suppression of star formation. While only a single example, this result shows that galaxy mergers can regulate the cold gas contents in distant galaxies by directly removing a large fraction of the molecular gas fuel, and plausibly suppress star formation directly, a qualitatively different physical mechanism than feedback-driven outflows.

\end{abstract}

\section{Introduction} \label{intro}

The regulation of galactic reservoirs of cold molecular gas -- the fuel for star formation -- over cosmic history is widely viewed as a critical but poorly-understood component of galaxy formation and evolution. A wide range of energetic `feedback' processes are required in modern galaxy formation simulations to moderate galaxy growth and produce galaxy populations that agree with even the most basic observational characteristics. Perhaps most strikingly, massive galaxy-scale outflows of cold molecular gas have been observed in galaxies and AGN at low and high redshifts with energetics capable of rapidly suppressing (or `quenching') ongoing star formation (\citealt{fluetsch19,spilker20b}, and see \citealt{veilleux20} for a recent review of cold gas outflows). Despite their undeniable importance in galaxy evolution, however, outflows driven by star formation and/or AGN need not be the only physical mechanisms relevant for all galaxy populations at all cosmic epochs.

Post-starburst galaxies, whose A-star dominated optical spectra indicate that an intense period of star formation was followed by rapid quenching $\approx$0.1--1\,Gyr ago, are valuable probes of star formation suppression as the direct and immediate products of quenching \citep[see][for a recent review]{french21}. While rare locally ($\sim$0.1\% of all galaxies), post-starbursts are 50$\times$ more common by $z\sim2$ \citep[e.g.][]{wild16} and represent about half of the growth of the quiescent population \citep[e.g.][]{belli19}. The burst in high-redshift post-starbursts also typically represents the primary formation phase of those galaxies, forming $\sim$70\% of the stellar mass during the burst compared to $\sim$10\% locally \citep{french18,wild20}. The \squiggle survey \citep{suess22} selected $\sim$1300 massive post-starburst galaxies from the SDSS at intermediate redshifts $z\sim0.7$, intended to be distant enough to be representative of the high-redshift post-starburst population yet nearby enough to allow detailed studies. Sample galaxies show no radial age gradients, implying quenching took place uniformly throughout the galaxies \citep{setton20}. Previous ALMA observations of a sample of \squiggle post-starbursts found that the quenching of star formation surprisingly does not require the total removal/destruction of the cold molecular gas reservoir, but this cold gas does rapidly disappear over the ensuing $\sim$100\,Myr post-quenching \citep{suess17,bezanson22}. 

Here we present the discovery of vast molecular tidal gas features reaching out from SDSS J144845.91+101010.5 (SDSS~J1448+1010), a $z=0.646$ post-starburst galaxy from the \squiggle survey. The cold gas contained in the tidal tails represents about half the cold gas in the entire system, and the nearly-concluded merger and accompanying tidal removal of the star-forming fuel is plausibly the proximate cause of the suppressed star formation in this galaxy. Section~\ref{data} describes the target galaxy and observational data, Section~\ref{results} presents our main findings, and we summarize and conclude in Section~\ref{conclusions}. We assume a \citet{chabrier03} initial mass function and a concordance flat $\Lambda$CDM cosmology with $\Omega_m=0.3$ and $h=0.7$.

\section{Data and Methods} \label{data}

\begin{figure*}
\begin{centering}
\includegraphics[width=\textwidth]{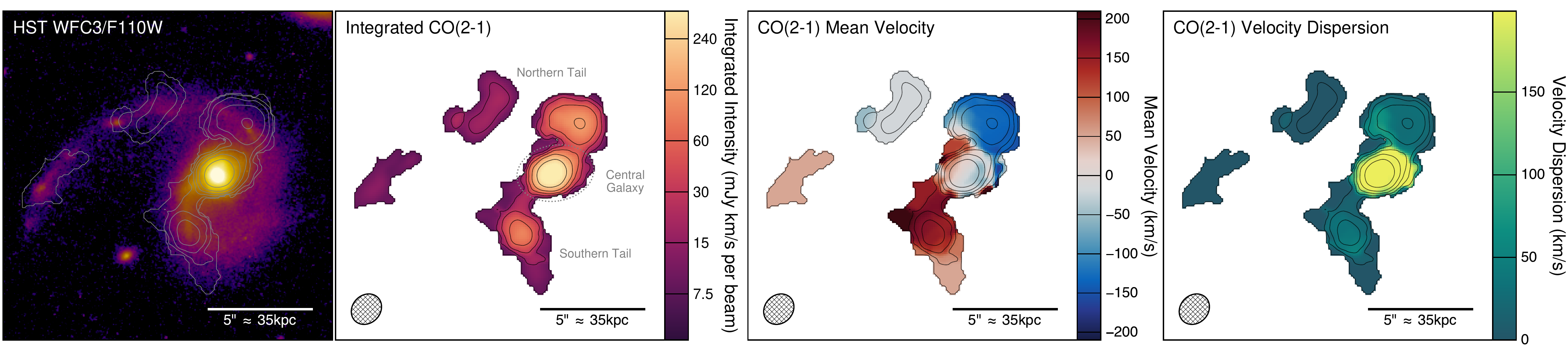}
\end{centering}
\caption{
Near-IR and CO emission of SDSS~J1448+1010.  CO is detected within the prominent stellar tidal tails seen in the \hst image, reaching projected distances up to 64\,kpc from the central galaxy. The northern (southern) tidal features are blue- (red-)shifted from the systemic velocity by $\approx$150\,\kms. The gas in the central galaxy is highly turbulent, while the dispersion is $<$50\,\kms at all locations within the tidal features. Contours of the integrated CO emission begin at 3$\sigma$ and increase in powers of 2, and are repeated in all panels. The ALMA synthesized beam is indicated at lower-left. The dotted circle in the second panel indicates the emission we consider to belong to the central galaxy; all else constitutes the tails.
}\label{fig:maps}
\end{figure*}

\subsection{ALMA Observations and Analysis} \label{almaobs}

SDSS~J1448+1010 was selected for CO(2--1) observations as part of a survey of 13 \squiggle post-starbursts presented in \citet{bezanson22}. The targets in \citet{bezanson22} sample the full range of spectral diversity of the parent sample (e.g. $D_{n}4000$, $H\delta_A$), but are preferentially brighter in the $i$ band due to a combination of lower than average redshift and higher than average stellar mass than the full sample. Initial ALMA Band~4 observations of SDSS~J1448+1010 were carried out in program 2017.1.01109.S targeting redshifted CO(2--1) at 140.06\,GHz. The total on-source time was 100\,min, sampling 15--780\,m baselines; see \citet{bezanson22} for further details. These data detected CO(2--1) emission extending well beyond the optical extent of the galaxy at $\approx5\sigma$ significance.

Concerned that large-scale low surface brightness emission may have been resolved out, we acquired deeper low-resolution data using the same observing setup in project 2018.1.01264.S. The new data sample 15--460\,m baselines, and the on-source integration time was 150\,min. We jointly imaged the data from both projects in CASA \citep{mcmullin07} using natural weighting to create both continuum images and CO(2--1) spectral cubes. We perform our subsequent analysis on a cube that averages 4 native channels with spectral resolution $\approx$33\,\kms. The final images reach a resolution of 1.3''$\times$1.6'', continuum sensitivity 4.5\,\uJy/beam, and spectral line sensitivity $\approx$97\,\uJy/beam in 33\,\kms channels.

\begin{figure*}
\centering{
\includegraphics[width=0.8\textwidth]{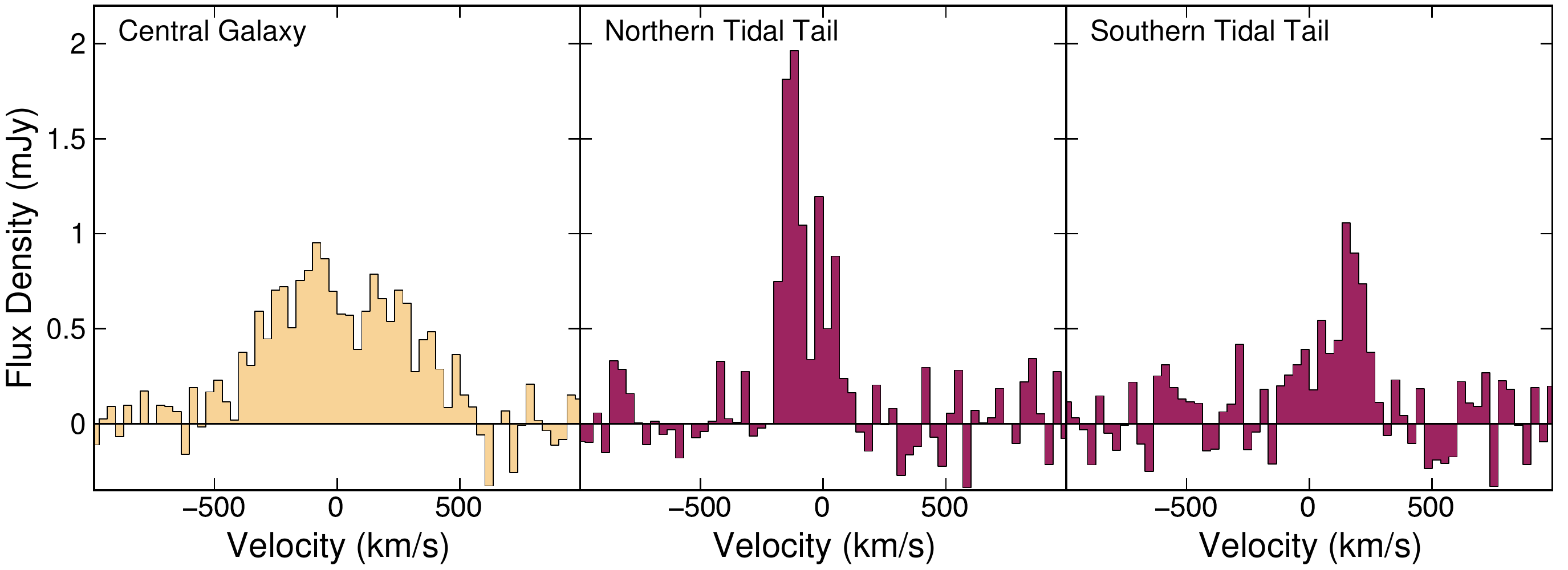}
}
\caption{
ALMA CO(2--1) spectra of SDSS~J1448+1010 extracted from, in order, the central galaxy and the northern and southern tidal tails. Combined, the tidal tails contain 47$\pm$5\% of the total CO luminosity in the system.
}\label{fig:spectra}
\end{figure*}

We use a slightly non-standard procedure to create CO(2--1) moment maps. Because the emission in the tidal features is much narrower than the central galaxy, simply integrating the CO cube over the spectral dimension would overwhelm the tidal features with many signal-free channels of noise. Instead we create moment maps by first identifying $\ge4\sigma$ peaks in the cube, then dilating outwards to the 2$\sigma$ contour in each spectral channel. The 4$\sigma$ threshold was chosen such that we expect $<1$ false peak within a 30''$\times$30''$\times$25 channel region of the cube given the synthesized beam size. Maps of the integrated CO emission, velocity field, and velocity dispersion are shown in Figure~\ref{fig:maps}. We extract CO spectra from within the dilated region, shown in Figure~\ref{fig:spectra}. We separate the emission from the galaxy from the northern and southern tidal tails by cutting perpendicular to the minimum level of the `valley' between the components in the integrated flux map, indicated by the dashed ellipse in Fig.~\ref{fig:maps}. We verified that this procedure produces integrated fluxes consistent with traditional aperture-based and image-plane fitting methods.

We detect weak continuum emission at the position of the central galaxy, $S_{\mathrm{147GHz}} = 19.3 \pm 4.5$\,\uJy. The flux ratio between the upper and lower sidebands (sky frequencies 153 and 141\,GHz), 0.9$\pm$0.3, is marginally inconsistent with that expected for dust emission on the Rayleigh-Jeans tail ($\approx$1.3--1.4 assuming a dust spectral index $\beta = 1.5-2$). This suggests that some of the 2\,mm continuum emission is due to non-thermal synchrotron emission instead of dust. Assuming a synchrotron spectral index $\alpha = -0.7$, we infer that $\gtrsim$25\% of the 2\,mm continuum emission is due to synchrotron emission.

\subsection{Near-Infrared Observations and Analysis} \label{nirobs}

\begin{figure*}
\centering{
\includegraphics[width=0.25\textwidth]{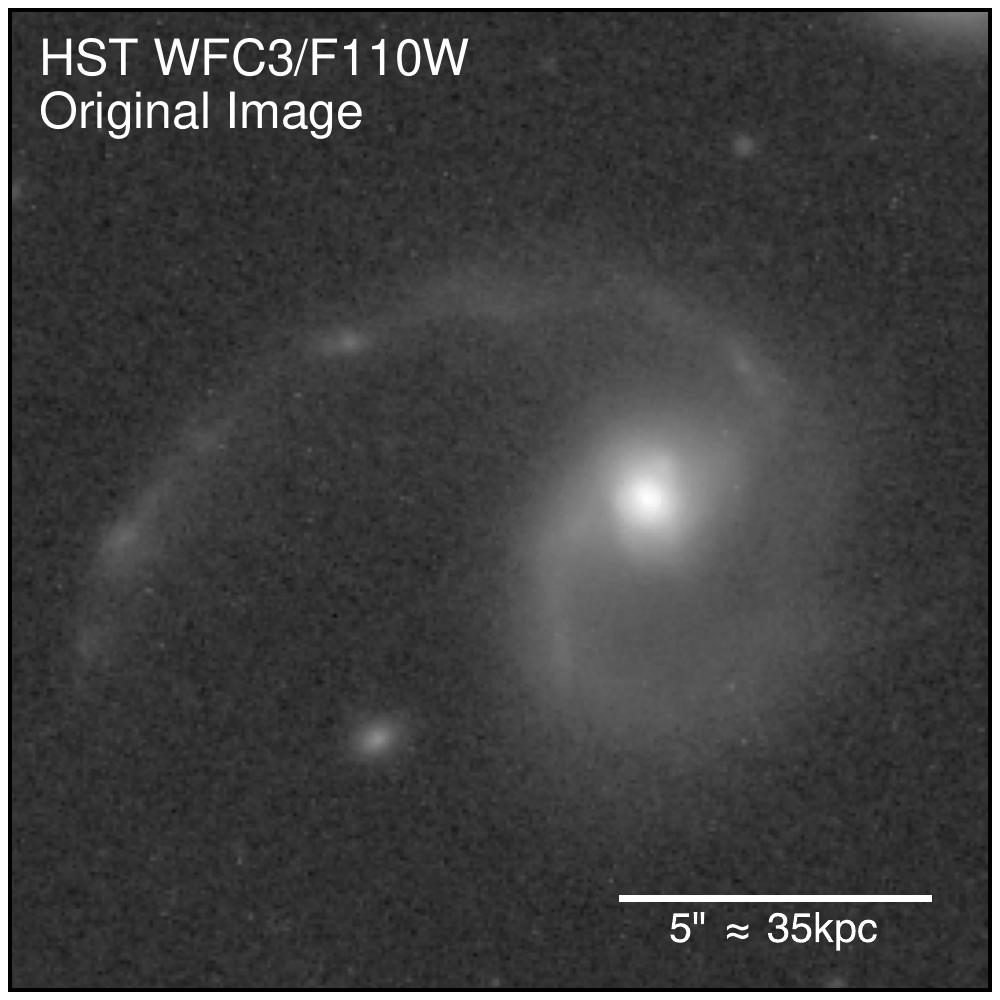}
\includegraphics[width=0.25\textwidth]{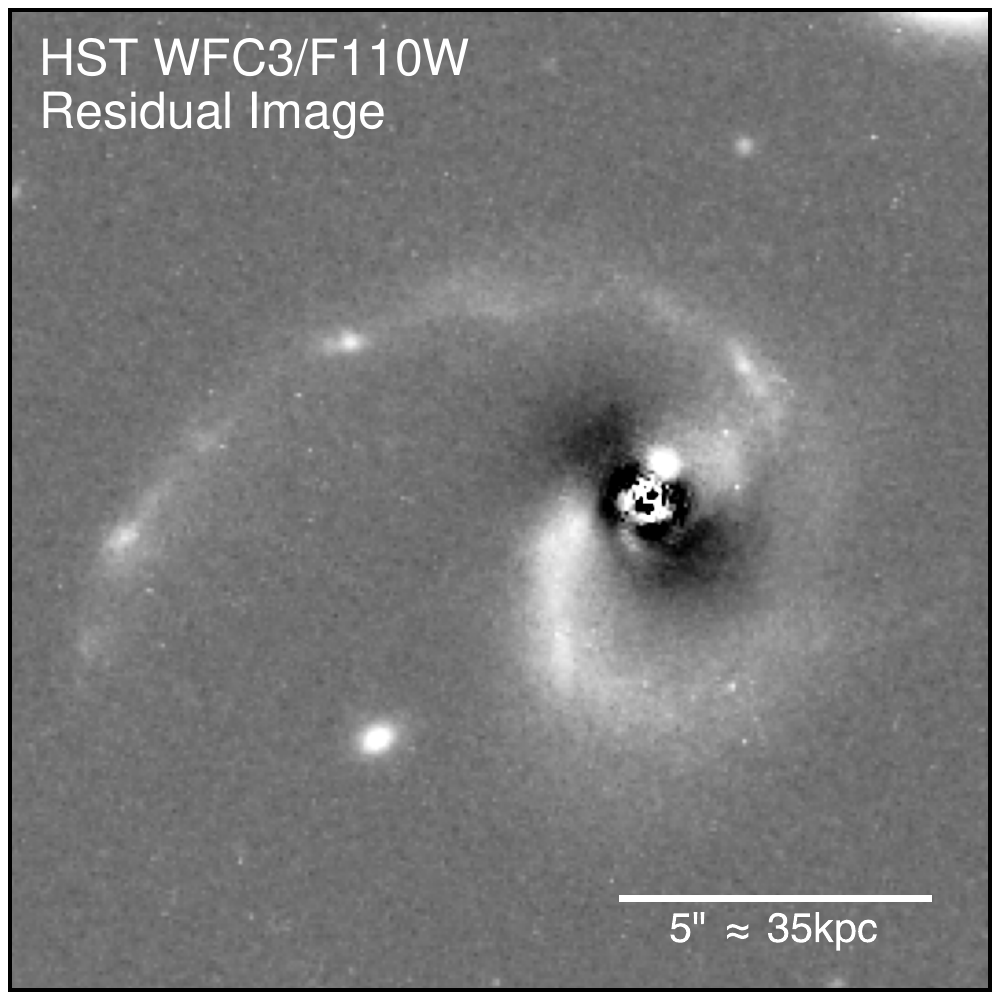}
\includegraphics[width=0.25\textwidth]{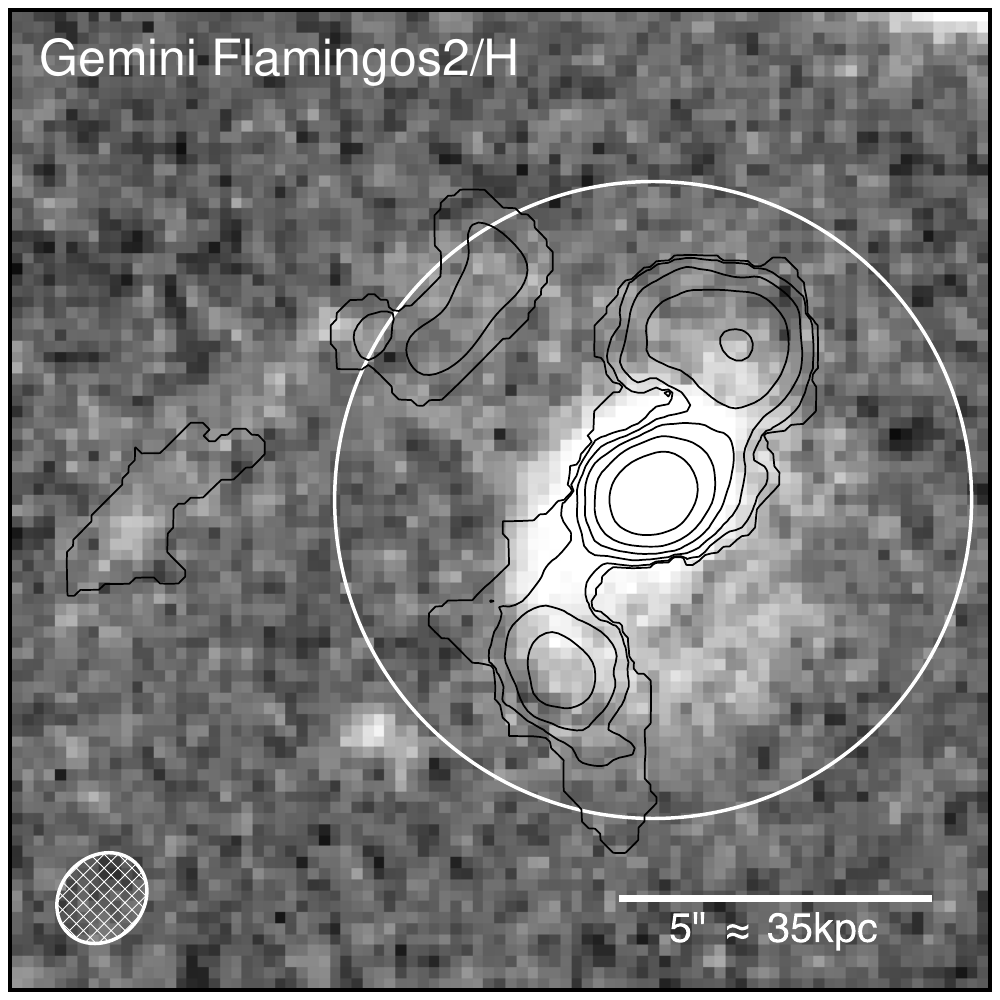}
}
\caption{
Left: \hst WFC3/F110W image of SDSS~J1448+1010, as in Fig.~\ref{fig:maps}, shown on log-stretch. Center: After subtraction of the best-fit single S\'{e}rsic profile, a second peak $\approx$0.6'' northwest of the main galaxy is more easily visible. Right: Gemini FLAMINGOS-2 H-band image with CO(2--1) contours. The white circle shows the photometric aperture used for all imaging data, matched to the SDSS catalog photometry. Note the object just off-frame to the northwest is an unrelated foreground galaxy.
}\label{fig:nirobs}
\end{figure*}

We observed SDSS~J1448+1010 in H-band using the FLAMINGOS-2 camera \citep{eikenberry04} on Gemini-South in program GS-2020A-FT-103 (PI Spilker). The instrument was used in imaging mode, and observations consisted of a repeating 9-point dither pattern with 8\,s exposures totaling 1440\,s on-source. The data were reduced using the Gemini \texttt{DRAGONS} software \citep{labrie19}, consisting of standard flat-field correction, dark and sky subtraction, alignment and stacking. The photometric calibration, checked against 2MASS stars in the field of view, is accurate to 0.04\,mag, and the seeing conditions average 0.5'' in the final image. The FLAMINGOS-2 image is shown in Figure~\ref{fig:nirobs}.

We also observed SDSS~J1448+1010 in a single orbit of \hst WFC3/IR F110W imaging in program PID 16201 (PI Spilker). The total exposure time was 2.45\,ks divided into a 4$\times$500\,s dither pattern and a 450\,s exposure at the final dither position. We fit a two-dimensional S\'{e}rsic profile to the galaxy using \texttt{galfit} \citep{peng10} after masking nearby sources and the tidal features, finding a best-fit half-light radius 0.54'' (3.8\,kpc). The size of the central galaxy agrees well with expectations for quiescent galaxies at this mass and epoch \citep{vanderwel14} and other \squiggle post-starbursts \citep{setton22}.

We identify a secondary peak near the center of the galaxy in the residual image (Fig.~\ref{fig:nirobs}); it is also visible by eye in the unsubtracted image with suitable stretch, and is not an artifact of the fitting procedure. We re-fit the image using a two-component model and found that the fainter component has very little impact on the best-fit parameters of the main galaxy. The peak is separated from the galaxy center by 0.6'' ($\approx$5 resolution elements), but more detailed properties are hampered by degeneracies due to the much brighter central peak. This secondary peak could be interpreted as a clump within the merger, a knot near the base of the northern tidal tail, or the remnant of the merging companion galaxy itself. In any case, the lack of an obvious merging companion implies that the merger is near final coalescence.

\subsection{Spectrophotometric Fitting} \label{sfhfit}

We jointly fit the available imaging and spectroscopic data following \citet{suess22} to measure the mass, SFR, and star formation history (SFH) of SDSS~J1448+1010. Briefly, we use the \texttt{prospector} SED fitting software \citep{leja17,johnson20} with a custom flexible SFH designed to be able to capture both a recent burst and variable level of star formation after the burst. This SFH has been extensively tested using mock data (Suess \etal in prep.). The starburst is defined as the time between which the derivative of the SFR rises above and declines below a threshold value \citep{suess22}. In addition to the SDSS photometry and spectroscopy and WISE photometry, we also include the photometric measurements from our new WFC3/F110W and FLAMINGOS-2 H-band imaging measured in a 5.2'' radius aperture matched to the SDSS catalogs. From the WFC3/F110W image, we estimate that an additional 6\% of the total light of the system is excluded by this aperture; this has negligible impact on our results.

\begin{figure*}
\centering{
\includegraphics[width=0.8\textwidth]{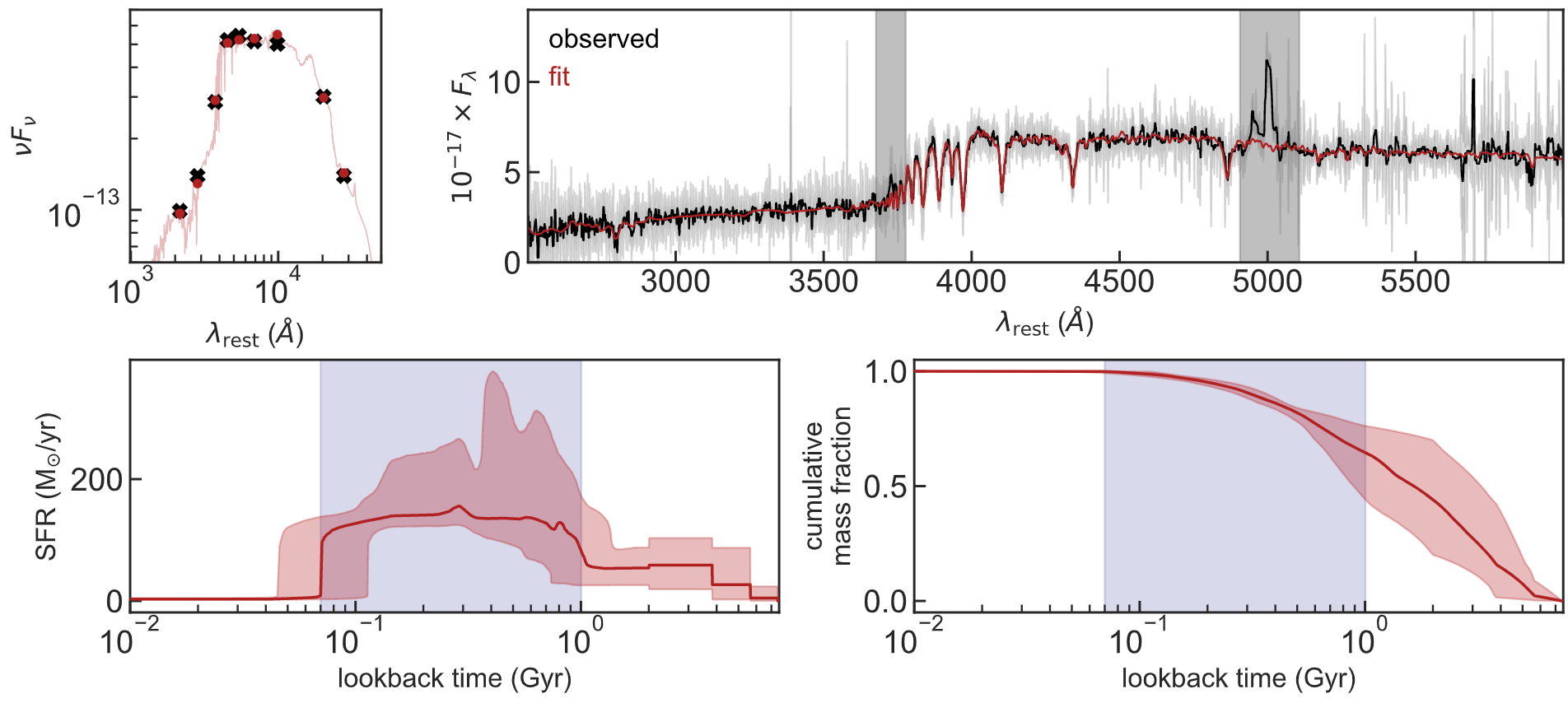}
}
\caption{
The available multiwavelength imaging and SDSS spectroscopic data (upper panels) allow detailed constraints on the star formation history of SDSS~J1448+1010. The optical spectrum shows prominent Balmer absorption features, the hallmark of post-starburst galaxies. The reconstructed SFH (lower panels) indicates that the starburst event began $1.0^{+0.3}_{-0.4}$\,Gyr and ended $70^{+40}_{-30}$\,Myr ago, and formed $\sim$40\% of the total stellar mass. The photometric data (black points) and best-fit model photometry (red points) are shown in the upper left panel. The SDSS spectrum (unbinned: gray; binned: black) and best-fit model spectrum (red) are shown in the upper right. Vertical gray bands highlight wavelengths around [OII]\,3727$\AA$ and [OIII]\,4959,5007$\AA$, which are masked in our fitting. Red shaded regions in the lower panels indicate the 16-84th percentile confidence interval on the reconstructed SFH, and the vertical blue shaded regions indicate the time period of the recent starburst.
}\label{fig:sfh}
\end{figure*}

The data, best-fit model, and reconstructed SFH are shown in Figure~\ref{fig:sfh}. We find a stellar mass $\log \Mstar/\Msol = 11.4\pm0.1$, metallicity $\log \Zstar/\Zsol = 0.2^{+0.2}_{-0.1}$, velocity dispersion $\sigma_{\mathrm{*}} = 210\pm13$\,\kms, and dust extinction $\mathrm{A}_\mathrm{V} = 0.4$\,mag. The SFH indicates that the galaxy had a starburst that began 1.0$^{+0.3}_{-0.4}$\,Gyr before the time of observation, ended 70$^{+40}_{-30}$\,Myr before observation, and produced 43$\pm$24\% of the galaxy's total stellar mass. After the conclusion of the starburst, the SFR dropped to $3^{+2}_{-1}$\,\Msol/yr by the observed epoch, a factor of 50--100 lower than the starburst peak.

It is possible this method has missed highly obscured star formation, extincted enough to be undetectable in the rest-UV to mid-IR data. We place a (loose) constraint on the obscured SFR using the ALMA 2\,mm detection; this is highly uncertain because we do not constrain the peak of the IR SED and the total IR luminosity depends strongly on the dust temperature \Tdust. 
Conservatively ignoring the possible $\gtrsim$25\% contribution to the 2\,mm flux density from synchrotron emission, we assume a modified blackbody with luminosity-weighted $\Tdust = 35$\,K and a standard conversion between \lir and SFR \citep{kennicutt12}. We find an obscured SFR\,$\sim10$\,\Msol/yr with large systematic uncertainty due to the unknown \Tdust. We note that \lir traces the SFR on timescales up to $\sim$100\,Myr \citep[e.g.][]{kennicutt12}, similar to the inferred time since quenching, which may explain why this estimate is higher than the `instantaneous' estimate from the spectral fitting. From the limited data available we find no evidence for substantial obscured star formation.

\section{Results and Discussion} \label{results}

\subsection{Vast and Massive Molecular Tidal Tails} \label{tailmass}

In addition to the central galaxy, we detect CO emission in two tail-like features extending up to $\approx$9'' east/northeast and south of the galaxy (Fig.~\ref{fig:maps}. At the redshift of SDSS~J1448+1010, these features are separated from the main galaxy by physical distances on the sky up to 64\,kpc. The northern features are mostly blueshifted by $\approx$100--150\,\kms and the southern feature redshifted by $\approx$100--150\,\kms. There is no discernible coherent velocity gradient in the gas within the central galaxy.\footnote{We also explored more aggressive visibility weighting schemes that improve resolution at the expense of sensitivity and found no evidence of ordered motion at $\approx$0.8'' resolution.} In contrast to the high velocity dispersion observed in the gas located within the central galaxy, $\sigma_\mathrm{CO} = 290 \pm 20$\,\kms, both extended features are narrow, with dispersion $<$50\,\kms at all locations. The gas within the central galaxy appears to be genuinely dispersion dominated -- it is implausible that inclination or resolution effects would result in such a high dispersion and lack of velocity gradient for a rotating system.

From the \hst/WFC3 and Gemini/FLAMINGOS-2 imaging (Fig.~\ref{fig:nirobs}), it is clear that the extended CO features are embedded within prominent stellar tidal features. Extended tidal tails such as these are produced in the late stages of galaxy-galaxy mergers with near-equal mass ratio between the merging partners \citep[e.g.][]{toomre72,cox08}. The spatial coincidence between the stellar and gas features indicates that the latter are also tidal in origin. In nearby galaxy mergers, tidal molecular gas is thought to have been stripped from the host galaxies in some cases, while in others the molecular gas may have formed in-situ from recycled atomic gas. In these latter cases, gas streams along the tidal tails and collects at the tips, where the pile-up increases the gas density and facilitates the conversion of some of the atomic gas into molecular form \citep[e.g.][]{braine00,bournaud04}. Because the molecular gas in the tidal tails of SDSS~J1448+1010 is not concentrated at the tips of the tidal tails, the hallmark of in-situ formation, we conclude that the cold gas was stripped from the pre-existing reservoirs of the galaxies by tidal forces as the merger proceeded.

CO spectra of the central galaxy and extended features are shown in Fig.~\ref{fig:spectra}. We fit the spectra with single Gaussian components; two Gaussians or directly summing over the channels with significant emission yielded consistent line fluxes. We find CO luminosities for the central galaxy and tails $\lpgal = (3.2 \pm 0.2) \times 10^9$\,K\,\kms\,pc$^2$ and $\lptail = (2.9 \pm 0.2) \times 10^9$\,K\,\kms\,pc$^2$, respectively. We assume standard literature conversions between CO luminosity and cold gas mass, namely thermalized CO emission and a CO-H$_2$ conversion factor $\alphaco = 4.0$\,\Msol/(K\,\kms\,pc$^2$), for both the central galaxy and the extended features, appropriate for non-starbursting, high-metallicty galaxies \citep[e.g.][]{bolatto13}. 

Under these assumptions we measure molecular masses $\Mmolgal = (1.3\pm0.1)\times10^{10}$\,\Msol for the galaxy and $\Mmoltail = (1.2\pm0.1)\times10^{10}$\,\Msol in the tails. Combined, the extended tidal tails contain 47$\pm$5\% of the total molecular gas in the system. We emphasize that this result derives from conservative assumptions. Estimates of the \textit{absolute} molecular masses of galaxy and tails depend on the absolute values of the CO(2--1)/CO(1--0) line ratio $r_{21}$ and \alphaco, while the \textit{fraction} of cold gas in the tails is only dependent on relative differences between the galaxy and tails. 

The CO excitation is a subdominant source of uncertainty due to the limited plausible range of $r_{21}$. Our assumption of thermalized emission ($r_{21} = 1$) is conservative, as lower values would imply higher gas masses. The value of \alphaco is more uncertain. For the central galaxy, we can cross-check our assumption using the 2\,mm continuum detection (although the tails are similarly massive, their larger spatial extent makes them undetected at the depth of our data). Although there is evidence that $>$25\% of the 2\,mm continuum arises from synchrotron emission (Sec.~\ref{almaobs}), we assume it is entirely due to dust emission to be conservative. Under standard assumptions for the dust emissivity and a mass-weighted dust temperature\footnote{Distinct from the luminosity-weighted \Tdust used to estimate \lir (Section~\ref{sfhfit}); see discussion in \citet{scoville16}.} of 25\,K \citep[e.g.][]{dunne03,scoville16}, we find a dust-based gas mass $\Mmolgal = (1.2\pm0.3)\times10^{10}$\,\Msol, implying $\alphaco \approx 3.8$, in excellent agreement with our assumptions. A $>$25\% contribution to the 2\,mm continuum from synchrotron emission would imply $\alphaco < 2.9$ for the central galaxy. If this limit on \alphaco applies to the galaxy but not the tails, the fraction of cold gas contained in the tidal tails would increase to $>$56\%.

\subsection{Extreme Cold Gas Ejection by a Major Merger} \label{lowzcomp}

The molecular gas properties of the tidal tails in SDSS~J1448+1010 are remarkable compared to all known galaxy mergers in the nearby or distant universe. We compiled a sample of 18 low-redshift ($z \lesssim 0.3$) galaxy mergers in which CO has been detected in the tidal tails \citep{combes88,becker89,smith99,heithausen00,schinnerer00,aalto01,braine00,braine01,braine03,braine04,duc07,lisenfeld08,boquien11,villarmartin13,sun14,sliwa17,fotopoulou19}. We also note the existence of two $z\sim1.5$ systems in which some of the detected CO emission was claimed to arise from tidally-dispersed material \citep{silverman18,puglisi21}. We excluded these objects from the comparison sample because the evidence that the emission is associated with tidal tails is indirect, based on subtraction or decomposition of low-resolution, low-S/N data. Consequently, the masses of host and tidal gas, the division between them, and the maximum spatial extent of any tidal gas are highly uncertain. The comparison sample is admittedly highly heterogeneous and spans a wide range of merger stage, total system mass, and so on. 

We systematized the literature sample to match our assumptions for the CO excitation and \alphaco. Compared to the original values, this increases the fraction of the gas in tidal tails for some objects (if the CO line opacity and therefore \alphaco was thought to be low in the tails) but lowers it for others (if a `starburst' \alphaco was assumed for the central galaxies). Uncertainties on the gas properties of the comparison sample, like for SDSS~J1448+1010, are dominated by the systematic uncertainty \alphaco. When not quoted by the original authors, we estimate the maximum projected separation of the tidal molecular features by measuring the extent from the galaxy nucleus to the most distant detected CO emission in the plane of the sky. These distances represent lower limits to the true maximum extent for all galaxies (including SDSS~J1448+1010) because the line-of-sight extent is unmeasured. As for SDSS~J1448+1010, we consider the molecular gas to have formed in situ if CO is detected at the tip of the tidal tail(s), or stripped otherwise \citep[e.g.][]{braine00,bournaud04}.

\begin{figure*}
\centering{
\includegraphics[width=0.9\textwidth]{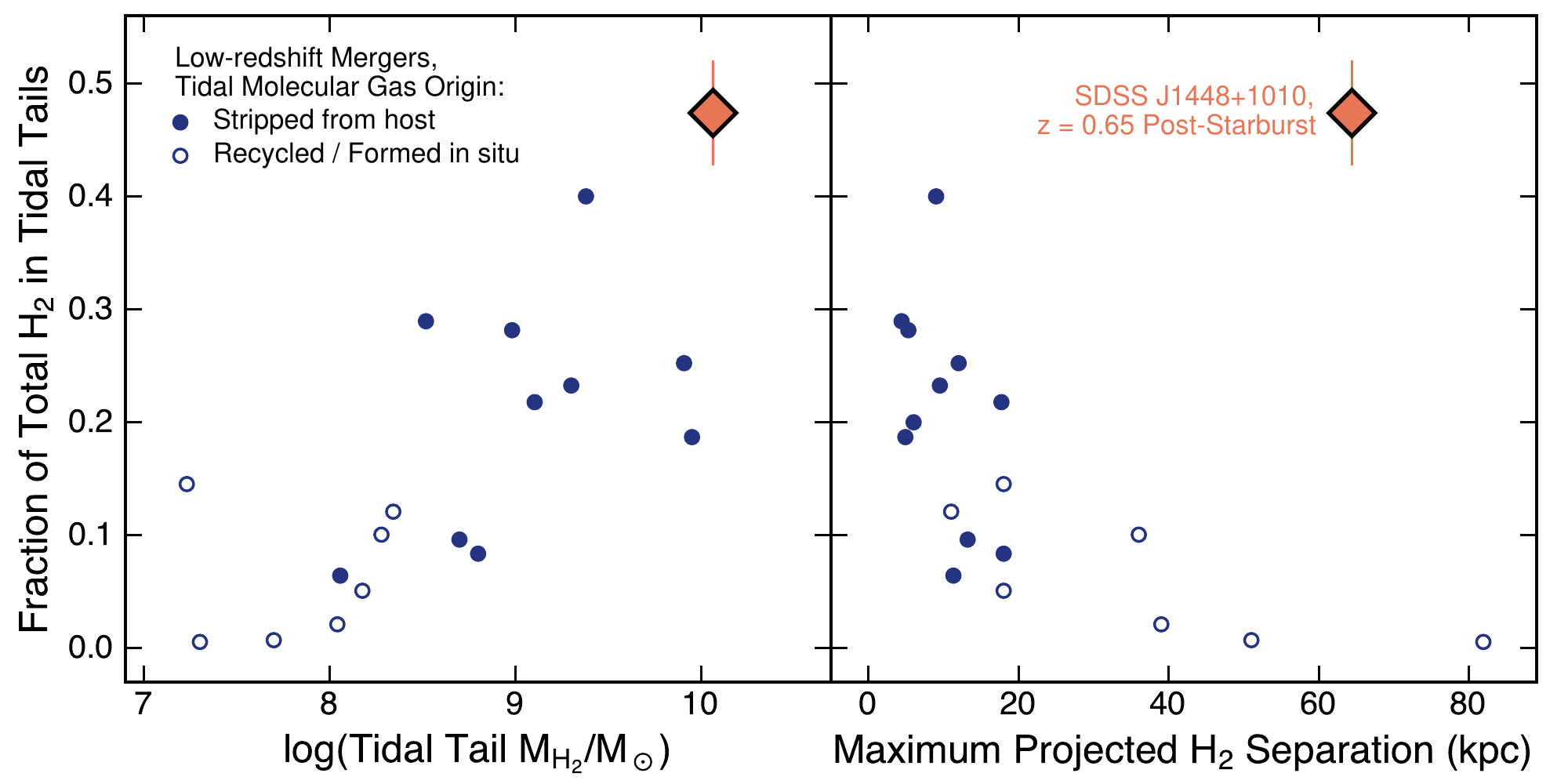}
}
\caption{
The molecular gas in the SDSS~J1448+1010 tidal tails are extreme in both scale and magnitude in comparison to all known mergers at any redshift. The low-redshift merger comparison sample is coded by whether or not the molecular gas is thought to have been stripped from the host (solid) or recycled / formed in situ (empty). All objects use the same assumptions for CO excitation and \alphaco; this figure would appear identical using pure observable quantities.
}\label{fig:lowzcomp}
\end{figure*}

Figure~\ref{fig:lowzcomp} compares the molecular gas properties of the tidal tails in SDSS~J1448+1010 with the CO detected in the nearby merger comparison sample. In terms of molecular gas contained in the tails, fraction of the total gas within the tails, and maximum extent of detected CO emission, SDSS~J1448+1010 is clearly extreme in comparison to nearby mergers. SDSS~J1448+1010 would also be highly unusual based on direct observables alone (i.e. CO luminosity). The molecular tidal features of nearby mergers broadly cluster in two groups depending on whether the molecular gas was formed in situ from atomic gas or stripped during the merger. Molecular gas can form in situ at large distances $\ge$80\,kpc, but this gas constitutes only a small fraction of the total cold gas contents of the system. Much larger amounts of molecular gas can be removed by tidal stripping, up to $\approx$40\% of the total, but in these cases the molecular features are only detected to projected distances $<$20\,kpc. SDSS~J1448+1010 is unique in comparison to these nearby mergers in both the scale and magnitude of the stripped molecular gas.

\subsection{Quenching by Tidal Cold Gas Removal} \label{quench}

Having shown that the molecular gas tidal tail properties of SDSS~J1448+1010 are extreme compared to all known examples in nearby galaxies, we finally briefly speculate on the origins and implications of these features for quenching the star formation in this massive post-starburst galaxy. We note up front that there are of course many physical processes that may have contributed to the quenching of star formation -- notably, this galaxy is also an optical AGN based on its \oiii/H$\beta$ line ratio, as in $\sim$10\% of the youngest post-starbursts in \squiggle \citep{greene20}. It is clear that the AGN did not produce the extended CO features, which perfectly trace the stellar tidal tails and have very cold dynamics, but it is less clear what role the AGN injection of thermal energy has contributed to the abrupt quenching or low ongoing SFR in this galaxy. Our goal here instead is to understand the potential consequences of the features thus far uniquely identified in SDSS~J1448+1010 -- namely, vast extended tidal tails of molecular gas.

While reconstructing the history of the merger is not possible with the available data, extended tidal tails such as those in SDSS~J1448+1010 are typical of the late stages of galaxy mergers. Analysis of the \textit{HST}/WFC3 image reveals a second peak in the near-IR light profile, separated from the galaxy center by $\sim$4.4\,kpc (Fig.~\ref{fig:nirobs}; Section~\ref{nirobs}). If interpreted as the remnant of the second galaxy in the merging pair, the small separation between nuclei also implies the merger is near final coalescence, $\sim$0.5--2\,Gyr after the interaction began \citep[e.g.][]{lacey93}. This timescale is consistent with the timing of the onset of the starburst found from the spectral modeling, estimated to have begun $1.0^{+0.3}_{-0.4}$\,Gyr prior to observation (Figure~\ref{fig:sfh}). Based on the well-constrained SFH of SDSS~J1448+1010, the most plausible physical scenario is that the initial stages of the merger also triggered the active starburst phase.

The tidal removal of molecular gas and the associated tidal stirring in the final mergers stages also may be directly responsible for the quenching of the starburst that concluded $70^{+40}_{-30}$\,Myr ago. By construction the \squiggle survey selects galaxies that experienced a sharp decline in SFR within the last billion years \citep{suess22}; a gradual decline in SFR cannot produce a galaxy-integrated spectrum with the prominent Balmer absorption features that are the hallmark of post-starburst galaxies. 

Tidal tails form and disperse over timescales of several hundred Myr \citep[e.g.][]{toomre72,hibbard95}, seemingly in conflict with the abrupt truncation of the starburst inferred from spectral modeling. However, the majority of the molecular gas in the tidal tails ($\sim$80\%, $\Mmol = 0.9\times10^{10}$\,\Msol) is located at projected galactocentric distances of 15--30\,kpc. From the CO velocity field (Fig.~\ref{fig:maps}) we see that this tidal material is traveling at a typical velocity $\approx$150\,\kms away from the central galaxy; this is a lower limit to the true gas velocity because motion in the plane of the sky is unobservable. At this minimum velocity the molecular gas in the tidal tails could have reached its present location in 100--200\,Myr, in good agreement with the timing of the end of the starburst. The cold gas not removed in the tidal interaction, meanwhile, has most likely been unable to form stars because it has been stirred into a highly turbulent state not conducive to star formation; the velocity dispersion of the gas in the galaxy, $\sigma_{\rm{CO}} = 290\pm20$\,\kms, is $\sim$50\% higher than the stellar velocity dispersion. Based on a larger sample of \squiggle post-starburst galaxies, we expect this cold gas to be destroyed, heated, and/or removed over the next $\sim$100\,Myr \citep{bezanson22}.

It is clear that the tidal removal of molecular gas from SDSS~J1448+1010 has had and will continue to have a large impact on the cold gas reservoir of this galaxy over a $>$1\,Gyr timescale, regardless of whether it is the proximate cause of the suppressed star formation. Galactic outflows of cold gas driven by supernovae and/or AGN are also often invoked to heat and disrupt the cold gas supply of galaxies and prevent the formation of over-massive galaxies \citep[e.g.][]{hopkins14,hayward17,dave19}. In comparison to molecular outflows detected at low and high redshifts, the tidal interaction in SDSS~J1448+1010 has removed a much larger fraction of the cold gas than typical outflows, which generally contain 1-10\% of the gas mass of the galaxy \citep[e.g.][]{fluetsch19,spilker20b}. In contrast, the velocity of the tidally-removed gas is likely much lower than the gas ejected in typical outflows, which reaches velocities of hundreds to $>$1000\,\kms \citep[e.g.][]{geach14,fluetsch19,spilker20a}. Tidal removal offers a slower, gentler mechanism by which the cold gas reservoir of galaxies is governed, played out over billion-year timescales.

\section{Synthesis and Conclusions} \label{conclusions}

We have discovered a spectacular vast molecular gas reservoir extending up to $>$60\,kpc outside of SDSS~J1448+1010, a $z=0.646$ massive post-starburst galaxy from the \squiggle survey. Near-IR imaging makes clear that the molecular gas features are co-located within extended tidal tails resulting from a recent/ongoing major merger. The tidal tails contain at least $47\pm5$\% of the total molecular gas in the system, and are extreme in both scale and magnitude compared to all known merging systems at any redshift. The timing of the merger and well-constrained star formation history are consistent with a scenario in which the initial phases of the merger triggered the recent starburst, while the subsequent rapid quenching event coincided with the ejection of most of the molecular gas into the tidal tails. 

From the single example we have discovered alone, it is unclear how common tidal ejection of cold gas from high-redshift galaxies may be. There are reasons to expect that tidal cold gas removal may be a much more common phenomenon at high redshift than today. High-redshift galaxies are far more gas-rich than modern galaxies, with the molecular mass increasing as $(1+z)^a$ with $a\approx 2.5-3$ \citep[e.g.][]{tacconi20}, which may explain why the tails in SDSS~J1448+1010 are so gas-rich relative to low-redshift mergers in absolute terms. The rate of galaxy mergers is also a strong function of cosmic age, increasing as $(1+z)^m$ with $m\approx2-3$ \citep[e.g.][]{lotz08,conselice09,bridge10,duncan19,ferreira20}, a consequence of the falling density of galaxies with time due to cosmic expansion. Together, the increased gas supply and higher merger rate suggest that mergers may play a far more important role in regulating the molecular gas contents of high-redshift galaxies than they do today. 

It is clear that the merger event has played and will continue to play a large role in the molecular gas contents of this system over long timescales. More speculatively, the tidal removal of the cold gas supply may have played a large role in directly quenching the star formation in this galaxy. Our discovery of vast and massive molecular tidal features was unexpected; future targeted observations will be able to determine how often this phenomenon occurs and place it in the broader context of galaxy quenching. This result opens up the possibility that tidal removal of the fuel for star formation may be a viable quenching mechanism that should be considered alongside galactic feedback and outflows.

\begin{acknowledgements}
JSS thanks Kathleen Labrie for assistance with the Gemini data reduction, and was supported in part by NASA Hubble Fellowship grant \#HF2-51446  awarded  by  the  Space  Telescope  Science  Institute,  which  is  operated  by  the  Association  of  Universities  for  Research  in  Astronomy,  Inc.,  for  NASA,  under  contract  NAS5-26555. This research used observations made with the NASA/ESA \textit{Hubble Space Telescope} operated under the same contract. DJS, RB, JEG, and DN acknowledge support from NSF-AAG \#1907697, \#1907723, and \#1908137.

This paper makes use of the following ALMA data: ADS/JAO.ALMA\#2017.1.01109.S, ADS/JAO.ALMA\#2018.1.01264.S. ALMA is a partnership of ESO (representing its member states), NSF (USA) and NINS (Japan), together with NRC (Canada), MOST and ASIAA (Taiwan), and KASI (Republic of Korea), in cooperation with the Republic of Chile. The Joint ALMA Observatory is operated by ESO, AUI/NRAO and NAOJ. The National Radio Astronomy Observatory is a facility of the National Science Foundation operated under cooperative agreement by Associated Universities, Inc.

Based on observations obtained at the international Gemini Observatory, a program of NSF's NOIRLab, which is managed by AURA under a cooperative agreement with the NSF on behalf of the Gemini Observatory partnership.
\end{acknowledgements}

\facility{ALMA, HST (WFC3), Gemini (FLAMINGOS-2), Sloan}

\software{
CASA \citep{mcmullin07},
\texttt{DRAGONS} \citep{labrie19},
\texttt{prospector} \citep{leja17,johnson20},
\texttt{astropy} \citep{astropy18},
\texttt{matplotlib} \citep{hunter07}}


\end{CJK*}
\end{document}